\documentclass[%
aps,
prb,
twocolumn
]{revtex4-2}

\usepackage[dvipdfmx]{graphicx}
\usepackage{mathtools}
\usepackage{amsmath}
\usepackage{amssymb}

\begin{document}


\title{Non-Hermitian skin effect in periodic, random, and quasiperiodic systems}


\author{F. Iwase}
\email[]{iwasef@tokyo-med.ac.jp}
\affiliation{Department of Physics, Tokyo Medical University\\
 6-1-1 Shinjuku, Shinjuku-ku, Tokyo 160-8402, Japan}



\begin{abstract}
The non-Hermitian skin effect (NHSE), which drives bulk states toward system boundaries, modifies bulk-boundary correspondence and complicates the identification of topological edge modes.
Although breaking translational symmetry is known to influence this behavior, a systematic comparison of different structural classes remains limited.
Here we investigate periodic, random, and quasiperiodic (Fibonacci) systems using a one-dimensional non-Hermitian quantum walk model.
By matching the local scattering parameters in a topologically nontrivial regime, we isolate the role of spatial structure in the presence of the NHSE.
We find that periodic systems exhibit pronounced boundary accumulation of bulk states.
Random systems suppress this accumulation through Anderson localization, but the topological gap becomes partially filled with localized in-gap states.
In contrast, the Fibonacci quasiperiodic system suppresses large-scale boundary accumulation while maintaining a well-defined topological gap.
Analysis of the wave functions suggests that the hierarchical quasiperiodic structure fragments bulk states across multiple length scales, thereby mitigating the NHSE.
These results identify deterministic quasiperiodicity as a distinct structural regime for controlling non-Hermitian skin dynamics and isolating topological boundary modes.
\end{abstract}


\maketitle

\section{Introduction\label{sec:intro}}

Non-Hermitian topological systems have attracted considerable interest in recent years, revealing phenomena that do not occur in Hermitian settings~\cite{Yao2018,Bergholtz2021,Ashida2020}.
A central feature of such systems is the non-Hermitian skin effect (NHSE), in which an extensive number of bulk states accumulate near system boundaries under open boundary conditions~\cite{Yao2018,Lee2016,Okuma2020,Zhang2022}.
This sensitivity to boundary conditions modifies the conventional bulk-boundary correspondence and strongly affects both the spatial structure of bulk states and the behavior of topological edge modes.

An important issue in this context is how the directional pumping associated with the NHSE interacts with mechanisms that inhibit the spatial spreading of bulk states in the absence of translational symmetry.
In disordered non-Hermitian systems, Anderson localization can compete with the skin effect, and sufficiently strong disorder suppress the boundary accumulation of bulk states by pinning the non-Hermitian directional flow~\cite{Hatano1996,Jiang2019,Claes2021}.
However, random disorder introduces a limitation for topological protection: it tends to generate localized in-gap states, such as Lifshitz tails, which populate the spectral gap and may hybridize with topological edge modes~\cite{Evers2008,Titum2015}.
As a result, the spectral isolation of the edge states can be reduced.

Deterministic quasiperiodic systems provide a different route to breaking the translational symmetry.
Sequences such as the Fibonacci lattice follow strict generation rules while lacking periodic order, and they are known to host fractal energy spectra and critically localized eigenstates~\cite{Kohmoto1983,Ostlund1983}.
These properties suggest that quasiperiodicity may influence the competition between non-Hermitian pumping and localization in a manner distinct from both periodic and randomly disordered systems.
Despite extensive studies of localization and topology in quasiperiodic lattices, a systematic comparison of how periodic, random, and quasiperiodic structures respond to the NHSE remains limited, particularly with respect to their spatial bulk dynamics and the preservation of topological gaps.

\begin{figure*}
\begin{center}
\includegraphics[width=16.5cm]{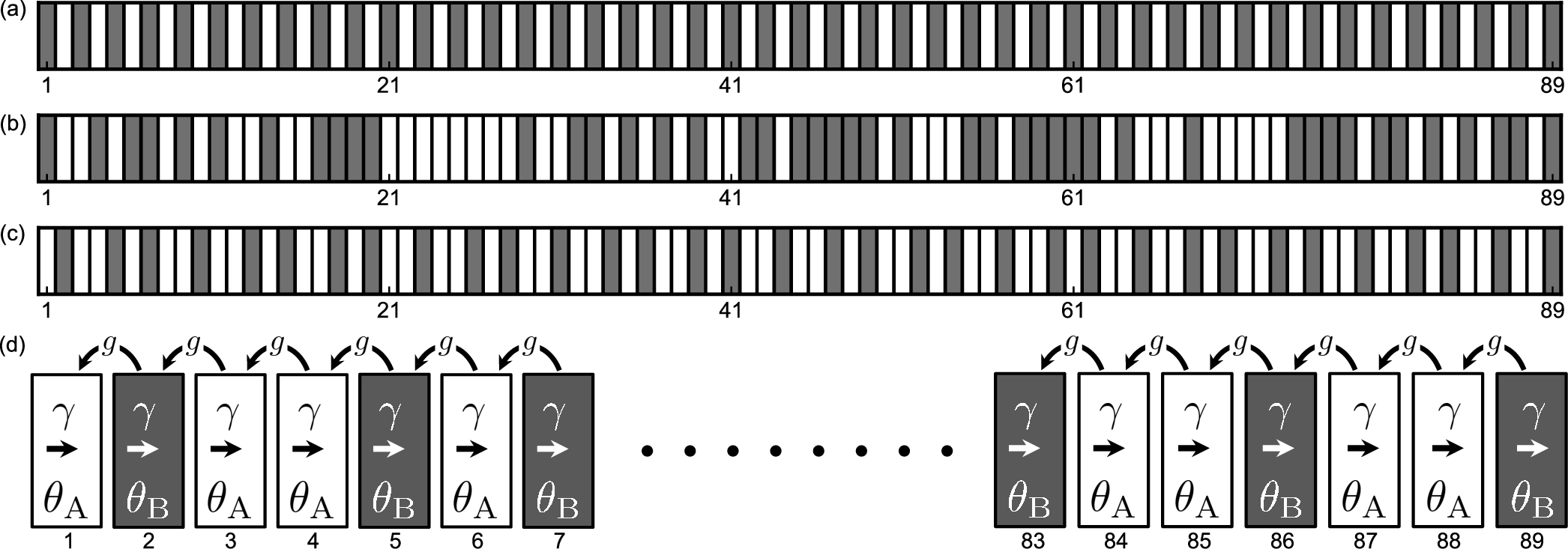}
\caption{\label{fig:schematic}
Schematic one-dimensional site sequences used in the three systems.
The system size is fixed at $N=89$.
White and gray blocks represent A and B sites, respectively, with coin parameters $\theta_\mathrm{A}=0.1\pi$ and $\theta_\mathrm{B}=0.51\pi$.
(a) Periodic, (b) Random, (c) Fibonacci quasiperiodic systems.
(d) Enlarged view of the left and right boundaries of the Fibonacci sequence.
The parameters $g$ and $\gamma$ represent non-Hermitian pumping toward the left and right directions, respectively.
}
\end{center}
\end{figure*}

In this work, the spatial and spectral responses of periodic, random, and Fibonacci quasiperiodic systems are investigated within a one-dimensional non-Hermitian quantum walk model.
To enable a controlled comparison, model parameters are chosen such that the three systems exhibit comparable scattering strengths and reside in the same topologically nontrivial phase in the Hermitian limit.
All systems are constructed from the same two internal coin degrees of freedom, while the spatial arrangement of these elements is varied to generate the periodic, random, and quasiperiodic structures.
In addition, the real-space topological phase at the right boundary, where the $\pi$ edge mode is localized, is verified to be identical in the three cases.
This construction enables a direct comparison in which the underlying local physics is identical and only the spatial order differs.

The results show that the periodic, random, and quasiperiodic structures exhibit qualitatively different responses to non-Hermitian skin effects.
In periodic systems the extended Bloch states undergo a strong boundary accumulation characteristic of the NHSE.
In random systems this accumulation is suppressed by Anderson localization~\cite{Anderson1958}, but the spectral gap becomes populated by localized impurity states.
In contrast, the Fibonacci quasiperiodic system suppresses large boundary accumulation while maintaining a comparatively clean topological gap.
These results clarify the distinct roles of periodic order, random disorder, and quasiperiodicity in shaping the response to the non-Hermitian skin effect.
In particular, they indicate that quasiperiodic order can suppress macroscopic boundary accumulation while maintaining spectral isolation of topological modes.

The remainder of this paper is organized as follows.
In Sec.~\ref{sec:method}, the non-Hermitian quantum walk model and the parameter choices used for the three systems are introduced.
In Sec.~\ref{sec:results1}, the quasienergy spectra in the Hermitian limit are analyzed to clarify the intrinsic spectral structures of the systems.
Section~\ref{sec:results2} examines their spatial dynamics through quantities such as center-of-mass shifts and inverse participation ratios.
In Sec.~\ref{sec:results3}, the complex spectra and spatial fragmentation of bulk states are discussed.
Section~\ref{sec:conclusion} summarizes the conclusions.

\section{Model and Formalism\label{sec:method}}
\subsection{Non-Hermitian quantum walk}

To systematically investigate the interplay between the NHSE and broken translational symmetries, we consider a discrete-time non-Hermitian quantum walk on a one-dimensional lattice~\cite{Asboth2012,Rudner2009,Mochizuki2016,Yao2018}.
The system consists of $N$ sites ($x\in\{0,\dots,N-1\}$) and the wave function resides in the composite Hilbert space $\mathcal{H}=\mathcal{H}_P\otimes \mathcal{H}_C$, where $\mathcal{H}_P$ is spanned by the position states $\{|x\rangle\}$ and $\mathcal{H}_C$ by coin states $\{|L\rangle, |R\rangle\}$.
The time evolution is generated by the non-unitary Floquet operator $U=SGC$.
Here, $C$ is the site-dependent coin operator implementing local rotations, while $G$ introduces non-Hermitian gain and loss:
\begin{eqnarray}
    C &=& \sum_{x=0}^{N-1} |x\rangle\langle x| \otimes 
    \begin{pmatrix} 
        \cos\theta_x & \sin\theta_x \\
        -\sin\theta_x & \cos\theta_x
    \end{pmatrix}, \\
    G &=& \sum_{x=0}^{N-1} |x\rangle\langle x| \otimes 
    \begin{pmatrix}
        e^{-\gamma} & 0 \\
        0 & e^\gamma
    \end{pmatrix}
\end{eqnarray}
The operator $S$ dictates the spatial shift.
To incorporate both the non-reciprocal pumping $g$ and the reflective boundary conditions at both ends of the lattice, $S$ is explicitly defined as:
\begin{multline}
    S = \sum_{x=1}^{N-1} e^g |x-1\rangle\langle x| \otimes |L\rangle\langle L| + \sum_{x=0}^{N-2} e^{-g} |x+1\rangle\langle x| \otimes |R\rangle\langle R| \\
    + |0\rangle\langle 0| \otimes |R\rangle\langle L| + |N-1\rangle\langle N-1| \otimes |L\rangle\langle R|.
\end{multline}

In this formulation, $\gamma > 0$ amplifies the right-moving component and thus induces an effective rightward pumping, whereas $g > 0$ introduces an asymmetric hopping amplitudes that favor leftward motion.
The competition between these two non-Hermitian mechanisms ($\gamma$ and $g$) determines the direction and magnitude of the resulting NHSE.

\subsection{Spatial sequences}

The structural configurations of the systems are introduced through the spatially dependent coin angles $\theta^{(x)}$.
We consider three types of sequences: periodic, random, and Fibonacci quasiperiodic.
All sequences are constructed from two constituent coin parameters, $\theta_\mathrm{A}$ and $\theta_\mathrm{B}$.

In the periodic system [Fig.~\ref{fig:schematic}(a)], the coin angles alternate as $\theta^{(x)} = \theta_\mathrm{B}, \theta_\mathrm{A}, \theta_\mathrm{B}, \theta_\mathrm{A}, \dots$, forming a perfectly periodic lattice.
For an odd system size $N$, the two boundaries therefore share the same coin type.
For clarity, the spatial coordinate is labeled as $x=1,\dots,N$ in the figure.

In the random system [Fig.~\ref{fig:schematic}(b)], the coin angles are assigned independently at each site as $\theta^{(x)}\in\{\theta_\mathrm{A}, \theta_\mathrm{B}\}$ with equal probability, representing a fully disordered structure. 

In the Fibonacci quasiperiodic system [Fig.~\ref{fig:schematic}(c)], the sequence is deterministically generated by the substitution rule A $\rightarrow$ AB, B $\rightarrow$ A, with the corresponding rotation angles $\theta_\mathrm{A}$ and $\theta_\mathrm{B}$ assigned to the symbols A and B, respectively.
The Fibonacci quasiperiodic sequence is known to generate a fractal (Cantor-like) spectrum with critically localized states~\cite{Kohmoto1983,Ostlund1983}, representing an intermediate order between periodic and random systems.

Figure~\ref{fig:schematic}(d) shows an enlarged view of the left and right boundaries of the Fibonacci sequence.
The non-Hermitian parameter $g$ is indicated by leftward arrows, while the site-dependent parameter $\gamma$ is represented by rightward arrows.

All systems considered in this work share the same system size $N=89$.
For the quasiperiodic case, this size corresponds to the Fibonacci sequence obtained after the 11th application of the substitution rule.
The periodic and random systems are constructed with the same size to enable a direct comparison among the three structural archetypes.
We confirmed that the qualitative behavior reported below remains unchanged for nearby system sizes.

\subsection{Parameter selection for controlled comparison}

A central requirement of the present study is that the comparison among the three systems reflects their global structural properties rather than trivial differences in local scattering strengths.
For this purpose, the same coin parameters are used for all systems, with $\theta_\mathrm{A}=0.1\pi$ and $\theta_\mathrm{B}=0.51\pi$.

In the periodic lattice, the boundary sites are chosen to have the coin $\theta_\mathrm{B}$ at both edges.
With this parameter choice, any differences in the response to the NHSE originate from the global structural characteristics of the lattices, namely extended Bloch states in periodic systems, Anderson localization in random systems, and critically localized states associated with the Cantor-like spectrum of the Fibonacci quasiperiodic structure.

For this configuration we verify that the zero mode is localized at the left boundary and the $\pi$ mode at the right boundary in all three systems.
The analysis presented below therefore focuses on the interaction between the bulk states and the right-edge $\pi$ mode characteristic of the Floquet system.

We have also confirmed the topological nontriviality of these parameter configurations in the Hermitian limit.
To treat the periodic and nonperiodic systems on an equal footing, we employed the real-space Schur-function analysis~\cite{Cedzich2018, Cedzich2022}, which yields an identical finite topological winding number $W=2$ for all three systems.
For the perfectly periodic system, we further verified that this result is consistent with the conventional bulk winding number $W=1$ obtained from the Zak phase~\cite{Zak1989, Asboth2012}.

These results confirm that all three systems reside in equivalent nontrivial topological phases and therefore support localized edge modes at the boundaries.
Consequently, any differences observed in the following analysis arise from the distinct spatial structures of the lattices rather than from differences in their topological protection.
This setup therefore provides a controlled framework for isolating the role of spatial structure in the emergence of the non-Hermitian skin effect.

\begin{figure}
\begin{center}
\includegraphics[width=8.5cm]{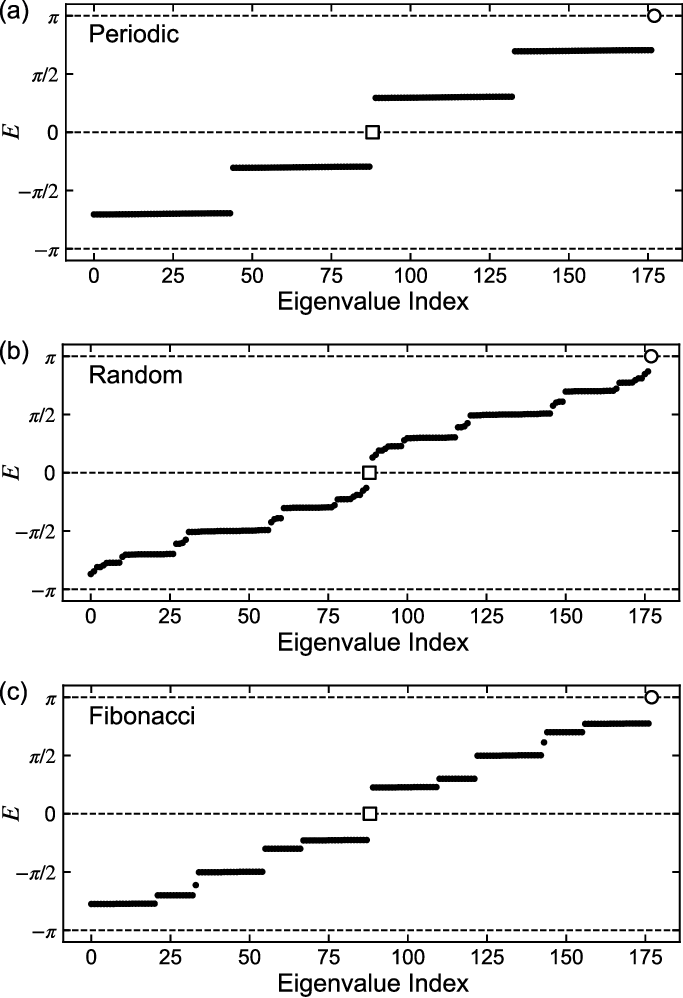}%
\caption{\label{fig:spectra}
Quasienergy spectra of periodic, random, and Fibonacci quasiperiodic systems in the Hermitian limit ($g=0$, $\gamma=0$).
(a) Periodic, (b) random, and (c) Fibonacci quasiperiodic spectra.
Eigenvalues are plotted as a function of the state index.
The isolated zero mode ($E=0$) and $\pi$ mode ($E=\pi$) are indicated by open squares and open circles, respectively.
}
\end{center}
\end{figure}

\section{Results\label{sec:results}}
\subsection{Intrinsic spectral structures in the Hermitian limit\label{sec:results1}}

Before introducing the non-Hermitian pumping effects, it is instructive to establish the intrinsic spectral properties of the three structural archetypes in the strictly Hermitian limit ($g=0$, $\gamma=0$).
Figure~\ref{fig:spectra}(a), (b), and (c) display the quasienergy $E$ (derived from the eigenvalues $\lambda=e^{-iE}$) as a function of the state index for the periodic, random, and Fibonacci systems, respectively.
Because our carefully chosen coin parameters ($\theta_\mathrm{A}=0.1\pi$, $\theta_\mathrm{B}=0.51\pi$ for all three systems) yield identical local scattering properties, all three systems exhibit a prominent topological gap around $E=0$ and $E=\pi$.
Within these gaps, the zero and $\pi$ edge modes are well resolved.

However, the bulk spectra surrounding these gaps reveal the fundamental differences.
In the perfectly periodic system, the quasienergy spectrum manifests as four macroscopic plateaus separated by wide band gaps.
These plateaus correspond to narrow bulk bands, a consequence of the strong local reflections induced by the chosen coin parameters.
Despite their narrow bandwidths, these states are inherently extended Bloch waves across the lattice.
In contrast, the random system exhibits characteristic Lifshitz tails; the sharp band edges are smeared, and the topological gap is partially filled by localized in-gap impurity states arising from Anderson localization.

Meanwhile, the Fibonacci quasiperiodic system displays a fractal energy spectrum, often described as a Cantor set~\cite{Kohmoto1983,Ostlund1983}.
Instead of the macroscopic plateaus seen in the periodic system, the quasiperiodic spectrum exhibits a hierarchical ``devil's staircase'' structure with multiple nested subgaps.
Despite the fragmentation of the bulk bands, the main topological gap hosting the edge modes remains relatively clean, with no significant in-gap states.

These intrinsically distinct spectral foundations---extended bands, disordered gap pollution, and fractal (Cantor-like) spectra---serve as the starting point for understanding their distinct dynamical responses under the NHSE.

\begin{figure}
\begin{center}
\includegraphics[width=8.5cm]{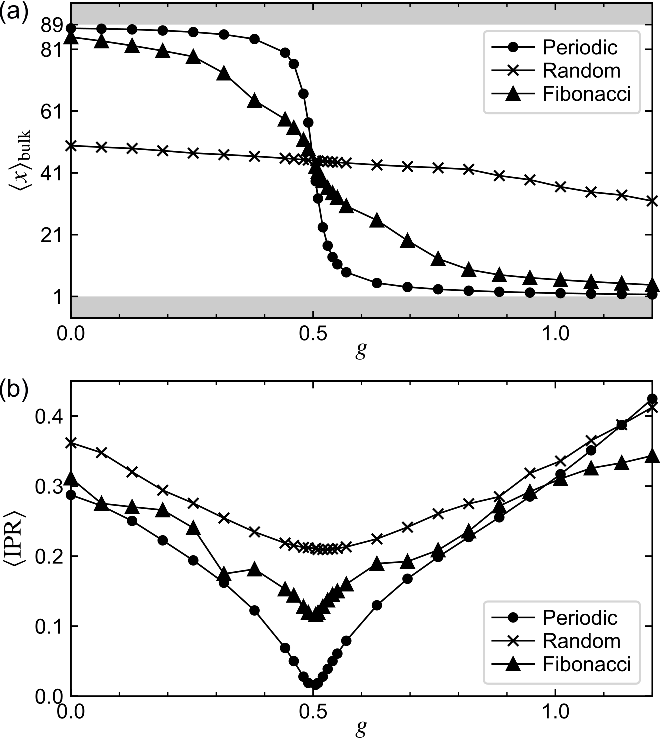}%
\caption{\label{fig:COM_and_IPR}
(a) Average center of mass (COM) and (b) inverse participation ratio (IPR) of the bulk states as functions of the nonreciprocal hopping parameter $g$.
Circles, squares, and triangles represent the periodic, random, and Fibonacci quasiperiodic systems, respectively.
The COM is measured in units of the lattice site index $x$.
The averages are taken over bulk eigenstates after excluding a few edge-localized states near both boundaries.
}
\end{center}
\end{figure}

\subsection{Spatial dynamics under non-Hermitian pumping\label{sec:results2}}

Having established the spectral properties in the Hermitian limit, we next examine the macroscopic spatial dynamics induced by the NHSE.
The local gain/loss parameter is fixed at $\gamma=0.5$, which enhances the right-moving components, while the non-reciprocal hopping parameter $g$ is varied to introduce a leftward pumping bias.
The net effective pumping direction is determined by the competition between $\gamma$ and $g$.
At the point $g=\gamma=0.5$, these opposing tendencies balance, yielding an effectively reciprocal Hamiltonian at the macroscopic level, in the sense that the net directional pumping vanishes.

\begin{figure*}
\begin{center}
\includegraphics[width=17.5cm]{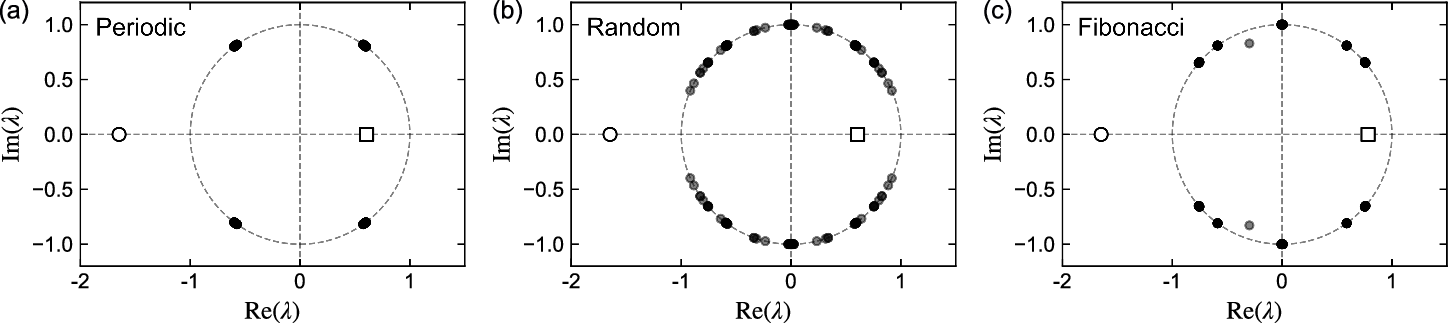}%
\caption{\label{fig:complex_spectrum}
Complex eigenvalue spectra of the three systems in the non-Hermitian regime ($g=0.4$, $\gamma=0.5$).
(a) Periodic, (b) random, and (c) Fibonacci quasiperiodic systems.
The two points on the real axis correspond to the $\pi$ (open circle) and zero modes (open square).
The $\pi$ mode lies outside the unit circle, indicating strong amplification.
}
\end{center}
\end{figure*}

To quantify the spatial distribution of the bulk states, we first calculate the average center of mass (COM), defined as 
\begin{equation}
    \langle x \rangle_\mathrm{bulk} = \frac{1}{N_\mathrm{bulk}}\sum_{i\in\mathrm{bulk}} \sum_{x=0}^{N-1} x |\psi_i(x)|^2,
\end{equation}
as a function of $g$ [Fig. \ref{fig:COM_and_IPR}(a)], where the site index $x$ in the figure is labeled from 1 to $N$.
Here the bulk set is defined by excluding edge-localized states near both boundaries, identified as those whose COM lies within the two sites of either edge.

In the periodic system the COM remains close to the right boundary and shows little dependence on $g$ for small $g$.
Around the critical value $g\approx 0.5$, however, it rapidly shifts toward the left boundary, indicating a strong non-Hermitian skin accumulation.
In contrast, the random system exhibits a much weaker response: the COM starts from a relatively small value and decreases gradually with increasing $g$, showing an almost linear dependence.
This behavior reflects the suppression of macroscopic drift by Anderson localization.
The Fibonacci quasiperiodic system displays an intermediate behavior.
Even for small $g$, the COM is not strongly localized near the right boundary, and it decreases gradually as $g$ increases.
The quasiperiodic system therefore suppresses the abrupt skin accumulation observed in the periodic case while still allowing a stronger drift than in the random system.


While the COM characterizes the spatial shift of the bulk wave packets, the inverse participation ratio (IPR) provides complementary information on their degree of spatial localization, defined as 
\begin{equation}
    \langle \mathrm{IPR} \rangle = \frac{1}{N_\mathrm{bulk}}\sum_{i\in\mathrm{bulk}} \sum_x |\psi_i(x)|^4.
\end{equation}
Figure~\ref{fig:COM_and_IPR}(b) shows the $\langle \mathrm{IPR} \rangle$ as a function of $g$.
The IPR curves exhibit a characteristic V-shaped profile centered at the cancellation point $g=0.5$, where the effective non-reciprocal pumping is canceled.
At this point the macroscopic NHSE is largely suppressed, allowing the intrinsic localization character of each system to be observed: the periodic system shows an IPR close to zero, consistent with extended Bloch states; the random system maintains a relatively large IPR reflecting Anderson localization; and the Fibonacci system exhibits an intermediate IPR, consistent with critically localized quasiperiodic states.

As $g$ deviates from $0.5$, the periodic system shows a rapid increase in the IPR, reflecting the transition from extended bulk states to boundary-accumulated skin modes.
For larger values of $g$, the periodic system eventually exhibits the largest IPR, reflecting the strong boundary accumulation associated with the NHSE.
The Fibonacci system, in contrast, exhibits a jagged V-shaped IPR.
This irregular behavior is consistent with hierarchical scattering induced by the quasiperiodic structure, where the non-Hermitian pumping progressively compresses the wave packets against spatially fragmented potential barriers.
As a result, the localization develops in a nonuniform and nonmonotonic manner, producing the structured IPR fluctuations observed in the quasiperiodic system.

Taken together, the COM and IPR analyses indicate that periodic system exhibit a collective boundary collapse of bulk states, random systems are dominated by Anderson-localized states, while the quasiperiodic system displays a fragmented bulk response intermediate between these two limits.

\subsection{Spectral and spatial structure of bulk states\label{sec:results3}}

To examine the microscopic behavior underlying the macroscopic trends discussed above, we focus on a representative parameter regime with $g=0.4$ and $\gamma=0.5$.
In this regime the effective non-Hermitian pumping remains rightward ($|\gamma | > |g|$), tending to drive bulk states toward the right boundary where the topological $\pi$-mode is localized at the right edge.

The complex spectra for the three systems in this regime are shown in Fig.~\ref{fig:complex_spectrum}.
In this representation, eigenvalues lying outside (inside) the unit circle correspond to amplified (decaying) modes.
Owing to the choice of coin parameters, all three systems support a strongly localized topological $\pi$-mode at the right boundary.
This corresponding eigenvalue lies well outside the unit circle with $|\lambda| \approx 1.65$, indicating a strong amplification of the $\pi$-mode.
Because the properties of the edge mode are thus nearly equalized, the essential differences among the systems arise from the behavior of the bulk states.

In the periodic system [Fig.~\ref{fig:complex_spectrum}(a)], the bulk eigenvalues form four nearly degenerate clusters in the complex plane, corresponding to the four plateaus observed in the quasienergy spectrum.
This structure reflects the underlying translational symmetry of the periodic lattice.
In the random system shown in Fig.~\ref{fig:complex_spectrum}(b), strong disorder introduces numerous localized in-gap eigenvalues that broadly populate the spectral gap through disorder-induced Lifshitz-tail states~\cite{Evers2008}, although a small local gap remains in the immediate vicinity of the $\pi$ modes. 
In contrast, the Fibonacci quasiperiodic system shown in Fig.~\ref{fig:complex_spectrum}(c) retains a clear spectral gap, with the $\pi$ mode remaining well isolated from the bulk spectrum even in the presence of non-Hermitian pumping.

\begin{figure}
\begin{center}
\includegraphics[width=8.5cm]{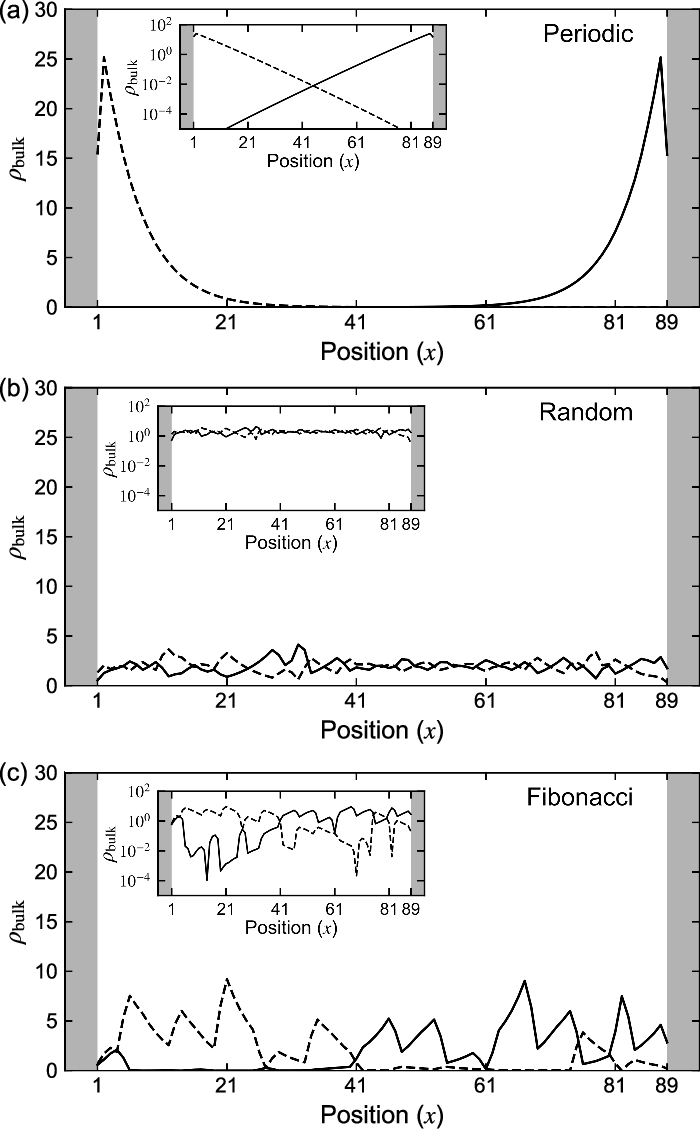}%
\caption{\label{fig:edge_bulk_density}
Spatial distribution of the total bulk density $\rho_\mathrm{bulk}(x)$ under non-Hermitian pumping ($\gamma=0.5$).
(a) Periodic, (b) random, and (c) Fibonacci quasiperiodic systems.
Solid and dashed curves correspond to $g=0.4$ and $g=0.6$, respectively.
The gray shaded area indicate the boundary areas excluded from the bulk region.
Insets show the logarithmic plots of $\rho_\mathrm{bulk}(x)$ down to $10^{-5}$.
$\rho_\mathrm{bulk}(x)$ is obtained by summing over eigenstates after excluding a few edge-localized states near both boundaries, identified by their center of mass within the two sites from each edge.
}
\end{center}
\end{figure}

This spectral purity must be evaluated together with the spatial distribution of the bulk states.
To quantify the spatial overlap between the bulk background and the topological edge mode, we compute the total bulk density defined as the sum of the spatial probabilities of all bulk eigenstates, 
\begin{equation}
    \rho_\mathrm{bulk}(x) = \sum_{i\in\mathrm{bulk}} |\psi_i(x)|^2.
\end{equation} 
The results are insensitive to the precise number of excluded edge-localized states.
Figure~\ref{fig:edge_bulk_density} shows $\rho_\mathrm{bulk}(x)$ under non-Hermitian pumping for the three systems.
Solid and dashed curves correspond to $g=0.4$ and $g=0.6$, respectively.

In the periodic system [Fig.~\ref{fig:edge_bulk_density}(a)], the bulk density accumulates near the right boundary for $g=0.4$ and near the left boundary for $g=0.6$.
This behavior reflects the fact that the extended Bloch states offer little structural resistance to the non-Hermitian pumping.
As a result, a large bulk background develops close to the boundary, leading to a strong spatial overlap with the $\pi$-mode or zero-mode.
As seen in the logarithmic inset, $\rho_\mathrm{bulk}(x)$ exhibits no internal spatial structure and decreases monotonically away from the boundary.

In the random system [Fig.~\ref{fig:edge_bulk_density}(b)], this macroscopic accumulation is suppressed by Anderson localization.
The localized bulk states appear at irregular positions across the lattice, producing a set of randomly distributed density peaks.

For the Fibonacci quasiperiodic system [Fig.~\ref{fig:edge_bulk_density}(c)], when $g=0.4$ the effective pumping drives states toward the right boundary, the bulk density becomes moderately biased toward the right side.
However, the strong boundary localization observed in the periodic system is largely suppressed.
Instead, the spatial distribution consists of several relatively broad localization peaks, forming a fragmented structure that differs both from the macroscopic accumulation in the periodic system and from the fine irregular fluctuations characteristic of the random system.
This behavior indicates that the hierarchical quasiperiodic structure scatters the bulk states across multiple length scales, leading to spatial fragmentation that prevents a macroscopic collapse of density at the boundary.

The logarithmic inset, shown down to $10^{-5}$, further reveals a small but finite bulk density even near the opposite edge, indicating that the bulk states remain spatially distributed over a broad region despite the directional pumping.
A similar situation occurs for $g=0.6$: although the overall density shifts toward the left side due to the reversed pumping direction, the boundary density remains small and the fragmented multi-peak structure is preserved.
This spatial fragmentation effectively reduces the overlap between the bulk background and the boundary-localized topological mode.
Consequently, the Fibonacci quasiperiodic structure simultaneously suppresses large-scale boundary accumulation of bulk states while maintaining a relatively clean spectral gap.

Although both random and quasiperiodic systems break translational symmetry, the underlying mechanisms are fundamentally different.
In random systems the suppression originates from Anderson localization accompanied by impurity states that populate the spectral gap, whereas in quasiperiodic systems the deterministic hierarchical potential produces fragmented bulk states while largely preserving a clean topological gap.

\section{Conclusion\label{sec:conclusion}}

In summary, we have investigated the macroscopic spatial dynamics and spectral properties of one-dimensional non-Hermitian topological systems under the NHSE.
By using a discrete-time non-Hermitian quantum walk, we performed a controlled comparison among periodic, random, and quasiperiodic (Fibonacci) systems.

The three structural archetypes exhibit qualitatively different responses to nonreciprocal pumping.
In periodic systems, the NHSE induces a macroscopic accumulation of bulk states at the boundary, leading to strong spatial overlap with the topological edge mode.
Random disorder suppresses this macroscopic accumulation through Anderson localization, but simultaneously introduces localized in-gap states that partially fill the spectral gap.

In contrast, the Fibonacci quasiperiodic system provides a distinct mechanism.
The deterministic hierarchical potential fragments the bulk states spatially and prevents large-scale boundary accumulation, while largely preserving a well-defined topological gap.
As a result, the topological edge mode remains well isolated from bulk states both spectrally and spatially.

These results highlight the role of deterministic aperiodicity as an effective structural strategy for controlling non-Hermitian skin dynamics.
Our findings suggest that quasiperiodic architectures may provide useful design principles for non-Hermitian topological devices, such as amplifiers, lasers, and sensors, where isolation of boundary modes from bulk noise is essential~\cite{Bandres2018,Helbig2020,Weidemann2020,Budich2020,Zou2021,Wanjura2020}.


\begin{thebibliography}{99}%
\bibitem{Yao2018} S. Yao, and Z. Wang, Edge States and Topological Invariants of Non-Hermitian Systems, Phys. Rev. Lett. \textbf{121}, 086803 (2018).
\bibitem{Bergholtz2021} E. J. Bergholtz, J. C. Budich, and F. K. Kunst, Exceptional topology of non-Hermitian systems, Rev. Mod. Phys. \textbf{93}, 015005 (2021).
\bibitem{Ashida2020} Y. Ashida, Z. Gong, and M. Ueda, Non-Hermitian Physics, Adv. Phys. \textbf{69}, 249 (2020).
\bibitem{Lee2016} T. E. Lee, Anomalous Edge State in a Non-Hermitian Lattice, Phys. Rev. Lett. \textbf{116}, 133903 (2016).
\bibitem{Okuma2020} N. Okuma, K. Kawabata, K. Shiozaki, and M. Sato, Topological Origin of Non-Hermitian Skin Effects
, Phys. Rev. Lett. \textbf{124}, 086801 (2020).
\bibitem{Zhang2022} X. Zhang, T. Zhang, M. Lu, and Y. Chen, A review on non-Hermitian skin effect, Adv. Phys. X \textbf{7}, 1 (2022).
\bibitem{Hatano1996} N. Hatano and D. R. Nelson, Localization Transitions in Non-Hermitian Quantum Mechanics, Phys. Rev. Lett. \textbf{77}, 570 (1996).
\bibitem{Jiang2019} H. Jiang, L. Lang, C. Yang, S. Zhu, and S. Chen, Interplay of non-Hermitian skin effects and Anderson localization in nonreciprocal low-dimensional systems, Phys. Rev. B \textbf{100}, 054301 (2019).
\bibitem{Claes2021} J. Claes, and T. L. Hughes, Skin effect and winding number in disordered non-Hermitian systems, Phys. Rev. B \textbf{103}, L014201 (2021).
\bibitem{Evers2008} F. Evers and A. D. Mirlin, Anderson transitions, Rev. Mod. Phys. \textbf{80}, 1355 (2008).
\bibitem{Titum2015} P. Titum, N. H. Lindner, M. C. Rechtsman, and G. Refael, Disorder-Induced Floquet Topological Insulators, Phys. Rev. Lett. \textbf{114}, 056801 (2015).
\bibitem{Kohmoto1983} M. Kohmoto, L. P. Kadanoff, and C. Tang, Localization Problem in One Dimension: Mapping and Escape, Phys. Rev. Lett. \textbf{50}, 1870 (1983).
\bibitem{Ostlund1983} S. Ostlund, R. Pandit, D. Rand, H. J. Schellnhuber, and E. D. Siggia, One-Dimensional Schrödinger Equation with an Almost Periodic Potential, Phys. Rev. Lett. \textbf{50}, 1873 (1983).
\bibitem{Anderson1958} P. W. Anderson, Absence of Diffusion in Certain Random Lattices, Phys. Rev. \textbf{109}, 1492 (1958).


\bibitem{Asboth2012} J. K. Asb\'oth, Symmetries, topological phases, and bound states in the one-dimensional quantum walk, Phys. Rev. B \textbf{86}, 195414 (2012).
\bibitem{Rudner2009} M. S. Rudner, and L. S. Levitov, Topological Transition in a Non-Hermitian Quantum Walk, Phys. Rev. Lett. \textbf{102}, 065703 (2009).
\bibitem{Mochizuki2016} K. Mochizuki, D. Kim, and H. Obuse, Explicit definition of $\mathcal{PT}$ symmetry for nonunitary quantum walks, Phys. Rev. A \textbf{93}, 062116 (2016).



\bibitem{Cedzich2018}
C. Cedzich, T. Geib, F. A. Gr\"{u}nbaum, C. Stahl, L. Vel\'{a}zquez, A. H. Werner, and R. F. Werner, The topological classification of One-Dimensional Symmetric Quantum Walks, Ann. Henri Poincar\'{e} \textbf{19}, 325 (2018).
\bibitem{Cedzich2022}
C. Cedzich, T. Geib, F. A. Gr\"{u}nbaum, L. Vel\'{a}zquez, A. H. Werner, and R. F. Werner, Quantum Walks: Schur Functions Meet Symmetry Protected Topological Phases, Commun. Math. Phys. \textbf{389}, 31 (2022).
\bibitem{Zak1989} J. Zak, Berry's phase for energy bands in solids, Phys. Rev. Lett. \textbf{62}, 2747 (1989).

\bibitem{Bandres2018} M. A. Bandres, S. Wittek, G. Harari, M. Parto, J. Ren, M. Segev, D. N. Christodoulides, and M. Khajavikhan, Topological insulator laser: Experiments, Science \textbf{359}, 1231 (2018).
\bibitem{Helbig2020} T. Helbig, T. Hofmann, S. Imhof, M. Abdelghany, T. Kiessling, L. W. Molenkamp, C. H. Lee, A. Szameit, and R. Thomale, Generalized bulk-boundary correspondence in non-Hermitian topolectrical circuits, Nat. Phys. \textbf{16}, 747 (2020).
\bibitem{Weidemann2020} S. Weidemann, M. Kremer, T. Helbig, T. Hofmann, A. Stegmaier, M. Greiter, R. Thomale, and A. Szameit, Topological funneling of light, Science \textbf{368}, 311 (2020).
\bibitem{Budich2020} J. C. Budich, and E. J. Bergholtz, Non-Hermitian Topological Sensors, Phys. Rev. Lett. \textbf{125}, 180403 (2020).
\bibitem{Zou2021} D. Zou, T. Chen, W. He, J. Bao, C. H. Lee, H. Sun, and X. Zhang, Observation of hybrid higher-order skin-topological effect in non-Hermitian topolectrical circuits, Nat. Commun. \textbf{12}, 7201 (2021).
\bibitem{Wanjura2020} C. C. Wanjura, M. Brunelli, and A. Nunnenkamp, Topological framework for directional amplification in driven-dissipative cavity arrays, Nat. Commun. \textbf{11}, 3149 (2020).








\end{thebibliography}
\end{document}